\documentclass[english]{article}
\usepackage[T1]{fontenc}
\usepackage[latin9]{inputenc}
\usepackage{amsmath}
\usepackage{graphicx}

\makeatletter
\usepackage{spconf}
\usepackage[pdftex,pdftitle=LPCNet: Improving Neural Speech Synthesis Through Linear Prediction,pdfauthor=Jean-Marc Valin]{hyperref}
\usepackage{hyphenat}
\makeatother

\usepackage{babel}
\begin{document}
\ninept
\twoauthors{Jean-Marc Valin}{Mozilla\\Mountain View, CA, USA\\\href{mailto:jmvalin@jmvalin.ca}{jmvalin@jmvalin.ca}}{Jan Skoglund}{Google LLC\\San Francisco, CA, USA\\\href{mailto:jks@google.com}{jks@google.com}}

\title{LPCNet: Improving Neural Speech Synthesis Through Linear Prediction}
\maketitle
\begin{abstract}
Neural speech synthesis models have recently demonstrated the ability
to synthesize high quality speech for text-to-speech and compression
applications. These new models often require powerful GPUs to achieve
real-time operation, so being able to reduce their complexity would
open the way for many new applications. We propose LPCNet, a WaveRNN
variant that combines linear prediction with recurrent neural networks
to significantly improve the efficiency of speech synthesis. We demonstrate
that LPCNet can achieve significantly higher quality than WaveRNN
for the same network size and that high quality LPCNet speech synthesis
is achievable with a complexity under 3~GFLOPS. This makes it easier
to deploy neural synthesis applications on lower-power devices, such
as embedded systems and mobile phones. 
\end{abstract}

\section{Introduction}

Neural speech synthesis algorithms have recently made it possible
to both synthesize high-quality speech~\cite{shen2018natural,arik2017deep,sotelo2017char2wav},
and code high quality speech at very low bitrate~\cite{kleijn2018wavenet}.
The first generation of these algorithms, often based on algorithms
like WaveNet~\cite{van2016wavenet}, gave promising results in real
time with a high-end GPU to provide the tens of billions of floating-point
operations per second (GFLOPS) required. We want to perform synthesis
on end-user devices like mobile phones, which do not have powerful
GPUs and have limited battery capacity.

Recent work~\cite{jin2018fftnet,kalchbrenner2018efficient} has focused
on finding more efficient models in order to reduce the complexity
of speech synthesis. In this work, we continue in that direction,
providing more efficiency improvements and making it easier to synthesize
speech even on slower CPUs, and with limited impact on battery life. 

Low complexity parametric synthesis models such as low bitrate vocoders
have existed for a long time~\cite{atal1971speech,markel1974linear},
but their quality has always been severely limited. While they are
generally efficient at modeling the spectral envelope (vocal tract
response) of the speech using linear prediction, no such simple model
exists for the excitation. Despite some advances~\cite{griffin1985new,mccree1996,juvela2018speaker},
modeling the excitation signal has remained a challenge. 

In this work, we propose the LPCNet model, which takes the burden
of spectral envelope modeling away from a neural synthesis network
so that most of its capacity can be used to model a spectrally flat
excitation. This makes it possible to match the quality of state-of-the
art neural synthesis systems with fewer neurons, significantly reducing
the complexity. Starting from the WaveRNN algorithm summarized in
Section~\ref{sec:WaveRNN}, we make improvements that reduce the
model complexity, as detailed in Section~\ref{sec:LPCNet}. In Section~\ref{sec:Evaluation},
we evaluate the quality and complexity of LPCNet in a speaker-independent
speech synthesis context based on the proposed model. We conclude
in Section~\ref{sec:Conclusion}.

\section{WaveRNN}

\label{sec:WaveRNN}

The WaveRNN architecture proposed in~\cite{kalchbrenner2018efficient}
takes as input the previous audio sample $s_{t-1}$, along with conditioning
parameters $\mathbf{f}$, and generates a discrete probability distribution
$P\left(s_{t}\right)$ for the output sample. Although it is proposed
as a 16-bit model (split as 8~coarse bits and 8~fine bits), we omit
the coarse/fine split in this summary, both for clarity and because
we do not use it in this work. The WaveRNN model mainly consists of
a gated recurrent unit (GRU)~\cite{cho2014properties}, followed
by two fully-connected layers, ending in a softmax activation. It
is computed as 
\begin{align}
\mathbf{x}_{t} & =\left[s_{t-1};\mathbf{f}\right]\nonumber \\
\mathbf{u}_{t} & =\sigma\left(\mathbf{W}^{\left(u\right)}\mathbf{h}_{t-1}+\mathbf{U}^{\left(u\right)}\mathbf{x}_{t}\right)\nonumber \\
\mathbf{r}_{t} & =\sigma\left(\mathbf{W}^{\left(r\right)}\mathbf{h}_{t-1}+\mathbf{U}^{\left(r\right)}\mathbf{x}_{t}\right)\nonumber \\
\widetilde{\mathbf{h}}_{t} & =\tanh\left(\mathbf{r}_{t}\circ\left(\mathbf{W}^{\left(h\right)}\mathbf{h}_{t-1}\right)+\mathbf{U}^{\left(h\right)}\mathbf{x}_{t}\right)\label{eq:WaveRNN}\\
\mathbf{h}_{t} & =\mathbf{u}_{t}\circ\mathbf{h}_{t-1}+\left(1-\mathbf{u}_{t}\right)\circ\widetilde{\mathbf{h}}_{t}\nonumber \\
P\left(s_{t}\right) & =\mathrm{softmax}\left(\mathbf{W}_{2}\ \mathrm{relu}\left(\mathbf{W}_{1}\mathbf{h}_{t}\right)\right)\ ,\nonumber 
\end{align}
where the $\mathbf{W}^{\left(\cdot\right)}$ and $\mathbf{U}^{\left(\cdot\right)}$
matrices are the GRU weights, $\sigma\left(x\right)=\frac{1}{1+e^{-x}}$
is the sigmoid function, and $\circ$ denotes an element-wise vector
multiply. Throughout this paper, biases are omitted for clarity. The
synthesized output sample $s_{t}$ is obtained by sampling from the
probability distribution $P\left(s_{t}\right)$. As a way of reducing
complexity, the matrices used by the GRU can be made sparse and~\cite{kalchbrenner2018efficient}
proposes using non-zero blocks of size 4x4 or 16x1 to ensure that
vectorization is still possible and efficient.

\section{LPCNet}

\label{sec:LPCNet}

\begin{figure}
\centering{\includegraphics[width=1\columnwidth]{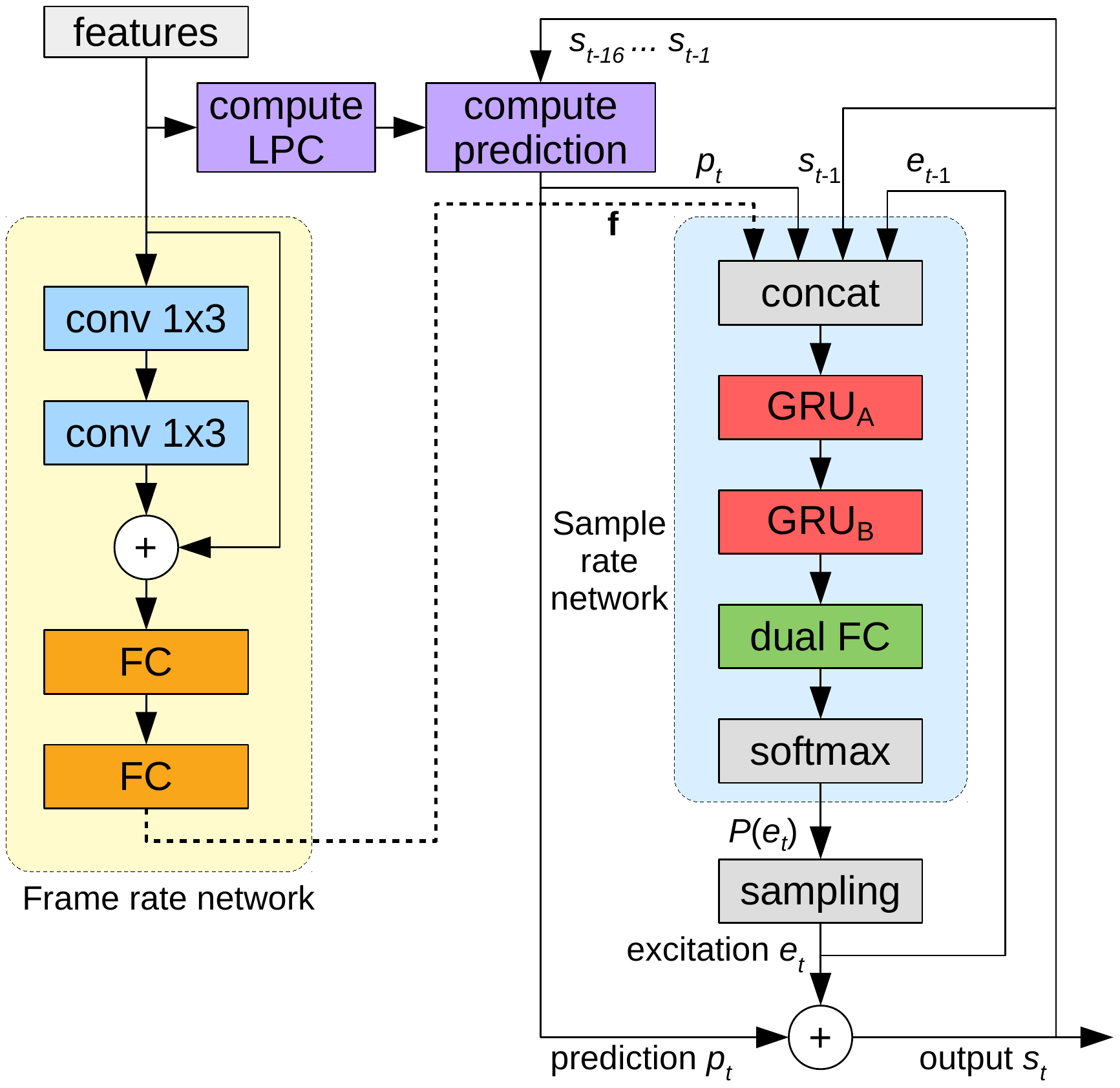}}

\caption{Overview of the LPCNet algorithm. The left part of the network (yellow)
is computed once per frame and its result is held constant throughout
the frame for the sample rate network on the right (blue). The \emph{compute
prediction} block predicts the sample at time $t$ based on previous
samples and on the linear prediction coefficients. Conversions between
$\mu$-law and linear are omitted for clarity. The de-emphasis filter
is applied to the output $s_{t}$.\label{fig:Overview-of-LPCNet} }
\end{figure}

This section presents the LPCNet model, our proposed improvement on
WaveRNN. Fig.~\ref{fig:Overview-of-LPCNet} shows an overview of
its architecture, which is explained in more details in this section.
It includes a sample rate network that operates at 16~kHz, and a
frame rate network that processes 10\nobreakdash-ms frames (160~samples).
In this work, we limit the input of the synthesis to just 20 features:
18~Bark-scale~\cite{moore2012introduction} cepstral coefficients,
and 2~pitch parameters (period, correlation). For low-bitrate coding
applications, the cepstrum and pitch parameters would be quantized~\cite{kleijn2018wavenet},
whereas for text-to-speech, they would be computed from the text using
another neural network~\cite{shen2018natural}. 

\subsection{Conditioning Parameters}

As part of the frame rate network, the 20 features first go through
two convolutional layers with a filter size of~3 (conv 3x1), resulting
in a receptive field of 5~frames (two frames ahead and two frames
back). The output of the two convolutional layers is added to a residual
connection and then goes through two fully-connected layers. The frame
rate network outputs a 128\nobreakdash-dimensional conditioning vector
$\mathbf{f}$ that is then used by the sample rate network. The vector
$\mathbf{f}$ is held constant for the duration of each frame.

\subsection{Pre-emphasis and Quantization}

Some synthesis models such as WaveNet~\cite{van2016wavenet} use
8-bit $\mu$-law quantization~\cite{G711} to reduce the number of
possible output sample values to just 256. Because the energy of speech
signals tends to be mostly concentrated in the low frequencies, the
$\mu$-law white quantization noise is often audible in the high frequencies,
especially for 16~kHz signals that have a higher spectral tilt. To
avoid this problem, some approaches extend the output to 16 bits~\cite{kalchbrenner2018efficient}.
Instead, we propose to simply apply a first-order pre-emphasis filter
$E(z)=1-\alpha z^{-1}$ to the training data, with $\alpha=0.85$
providing good results. The synthesis output can then be filtered
using the inverse (de-empahsis) filter 
\begin{equation}
D(z)=\frac{1}{1-\alpha z^{-1}}\ ,\label{eq:deemphasis}
\end{equation}
effectively shaping the noise such that its power at the Nyquist rate
is reduced by 16~dB. This significantly reduces the perceived noise
(see Section~\ref{subsec:Quality}) and makes 8-bit $\mu$-law output
viable for high-quality synthesis.

\subsection{Linear Prediction}

The neural networks in many neural speech synthesis approaches~\cite{kleijn2018wavenet,van2016wavenet,jin2018fftnet,kalchbrenner2018efficient,mehri2016samplernn}
have to model the entire speech production process, including glottal
pulses, noise excitation, as well as the response of the vocal tract.
Although some of these are indeed very hard to model, we know that
the vocal tract response can be represented reasonably well by a simple
all-pole linear filter~\cite{makhoul1975linear}. Let $s_{t}$ be
the signal at time $t$, its linear prediction based on previous samples
is
\begin{equation}
p_{t}=\sum_{k=1}^{M}a_{k}s_{t-k}\ ,\label{eq:prediction}
\end{equation}
where $a_{k}$ are the $M^{th}$ order linear prediction coefficients
(LPC) for the current frame. 

The prediction coefficients $a_{k}$ are computed by first converting
the 18\nobreakdash-band Bark-frequency cepstrum into a linear-frequency
power spectral density (PSD). The PSD is then converted to an auto-correlation
using an inverse FFT. From the auto-correlation, the Levinson-Durbin
algorithm is used to compute the predictor. Computing the predictor
from the cepstrum ensures that no additional information needs to
be transmitted (in a speech coding context) or synthesized (in a text-to-speech
context). Even though the LPC analysis computed in this way is not
as accurate as one computed on the input signal (due to the low resolution
of the cepstrum), the effect on the output is small because the network
is able to learn to compensate. This is an advantage over \emph{open-loop}
filtering approaches~\cite{juvela2018speaker}. 

As an obvious extension of using a linear predictor to help the neural
network, we can also have the network directly predict the excitation
(prediction residual) rather than the sample values. This is not only
a slightly easier task for the network, but it also slightly reduces
the $\mu$-law quantization noise, since the excitation generally
has a smaller amplitude than the pre-emphasized signal. The network
takes as input the previously sampled excitation $e_{t-1}$, but also
the past signal $s_{t-1}$, and the current prediction $p_{t}$. We
still include $s_{t-1}$ and $p_{t}$ because we find that \emph{open-loop}
synthesis based only on $e_{t-1}$ produces bad quality speech (see
Section~\ref{subsec:training-noise-injection}).

\subsection{Output Layer}

To make it easier to compute the output probabilities without significantly
increasing the size of the preceding layer, we combine two fully-connected
layers with an element-wise weighted sum. The layer, which we refer
to as \emph{dual fully-connected} (or DualFC) is defined as
\begin{equation}
\mathrm{dual\_fc}(\mathbf{x})=\mathbf{a}_{1}\circ\tanh\left(\mathbf{W}_{1}\mathbf{x}\right)+\mathbf{a}_{2}\circ\tanh\left(\mathbf{W}_{2}\mathbf{x}\right)\ ,\label{eq:dual_fc}
\end{equation}
where $\mathbf{W}_{1}$ and $\mathbf{W}_{2}$ are weight matrices,
and $\mathbf{a}_{1}$ and $\mathbf{a}_{2}$ are weighting vectors.
While not strictly necessary for the proposed approach to work, we
have found that the DualFC layer slightly improves quality when comparing
to a regular fully-connected layer at equivalent complexity. The intuition
behind the DualFC later is that determining whether a value falls
withing a certain range ($\mu$-law quantization interval in this
case) requires two comparisons, with each fully-connected $\mathrm{tanh}$
layer implementing the equivalent of one comparison. Visualizing the
weights on a trained network supports that intuition. The output of
the DualFC layer is used with a softmax activation to compute the
probability $P\left(e_{t}\right)$ of each possible excitation value
for $e_{t}$.

\subsection{Sparse Matrices}

To keep the complexity low, we use sparse matrices for the largest
GRU ($\mathrm{GRU_{A}}$ in Fig.~\ref{fig:Overview-of-LPCNet}).
Rather than allowing general element-by-element sparseness -- which
prevents efficient vectorization -- we use block-sparse matrices
as proposed in~\cite{kalchbrenner2018efficient}. Training starts
with dense matrices and the blocks with the lowest magnitudes are
progressively forced to zero until the desired sparseness is achieved.
We find that 16x1~blocks provide good accuracy, while making it easy
to vectorize the products. 

In addition to the non-zero blocks, we also include all the diagonal
terms in the sparse matrix, since those are the most likely to be
non-zero. Even though they are not aligned horizontally or vertically,
the diagonal terms are nonetheless easy to vectorize since they result
in an element-wise multiplication with the vector operand. Including
the diagonal terms avoids forcing 16x1~non-zero blocks only for a
single element on the diagonal.

\subsection{Embedding and Algebraic Simplifications}

Rather than scale the scalar sample values to a fixed range before
feeding them to the network, we use the discrete nature of the $\mu$-law
values to learn an embedding matrix $\mathbf{E}$. The embedding maps
each $\mu$-law level to a vector, essentially learning a set of non-linear
functions to be applied to the $\mu$-law value. Visualizing the embedding
matrix of trained networks, we have been able to confirm that the
embedding has learned -- among other things -- the function that
converts the $\mu$-law scale to linear. 

As long as the embedding is sent directly to the GRU, it is possible
to avoid increasing the complexity by pre-computing the product of
the embedding matrices with the corresponding submatrices of the GRU's
non-recurrent weights ($\mathbf{U}^{\left(\cdot\right)}$). Let $\mathbf{U}^{\left(u,s\right)}$
be the submatrix of $\mathbf{U}^{\left(u\right)}$ composed of the
columns that apply to the embedding of the $s_{t-1}$ input sample,
we can derive a new embedding matrix $\mathbf{V}^{\left(u,s\right)}=\mathbf{U}^{\left(u,s\right)}\mathbf{E}$
that directly maps the sample $s_{t-1}$ to the non-recurrent term
of the update gate computation. The same transformation applies for
all gates ($u$, $r$, $h$) and all embedded inputs ($s$, $p$,
$e$), for a total of 9~pre-computed $\mathbf{V}^{\left(\cdot,\cdot\right)}$
matrices. In that way, the embedding contribution can be simplified
to only one add operation per gate, per embedded input. Since only
a single entry per embedding matrix is used for each sample, the large
size of these matrices is not an issue, even if they do not fit in
cache. 

In a similar way to the embedding, the contributions of the frame
conditioning vector $\mathbf{f}$ can also be simplified since it
is constant over an entire frame. Once per frame, we can compute $\mathbf{g}^{\left(\cdot\right)}=\mathbf{U}^{\left(\cdot\right)}\mathbf{f}$,
the contribution of $\mathbf{f}$ to each of the GRU gates. 

The simplifications above essentially make the computational cost
of all the non-recurrent inputs to the main GRU negligible, so the
calculations in~(\ref{eq:WaveRNN}) become
\begin{align}
\mathbf{u}_{t}= & \sigma\left(\mathbf{W}_{u}\mathbf{h}_{t}+\mathbf{v}_{s_{t-1}}^{\left(u,s\right)}+\mathbf{v}_{p_{t-1}}^{\left(u,p\right)}+\mathbf{v}_{e_{t-1}}^{\left(u,e\right)}+\mathbf{g}^{\left(u\right)}\right)\nonumber \\
\mathbf{r}_{t}= & \sigma\left(\mathbf{W}_{r}\mathbf{h}_{t}+\mathbf{v}_{s_{t-1}}^{\left(r,s\right)}+\mathbf{v}_{p_{t-1}}^{\left(r,p\right)}+\mathbf{v}_{e_{t-1}}^{\left(r,e\right)}+\mathbf{g}^{\left(r\right)}\right)\label{eq:LPCNet}\\
\widetilde{\mathbf{h}}_{t}= & \tanh\left(\mathbf{r}_{t}\circ\left(\mathbf{W}_{h}\mathbf{h}_{t}\right)+\mathbf{v}_{s_{t-1}}^{\left(h,s\right)}+\mathbf{v}_{p_{t-1}}^{\left(h,p\right)}+\mathbf{v}_{e_{t-1}}^{\left(h,e\right)}+\mathbf{g}^{\left(h\right)}\right)\nonumber \\
\mathbf{h}_{t}= & \mathbf{u}_{t}\circ\mathbf{h}_{t-1}+\left(1-\mathbf{u}_{t}\right)\circ\widetilde{\mathbf{h}}_{t}\nonumber \\
P\left(e_{t}\right) & =\mathrm{softmax}\left(\mathrm{dual\_fc}\left(\mathrm{GRU_{B}}\left(\mathbf{h}_{t}\right)\right)\right)\ ,\nonumber 
\end{align}
where the $\mathbf{v}_{i}^{\left(\cdot,\cdot\right)}$ vectors are
lookups of column vector $i$ into the corresponding $\mathbf{V}^{\left(\cdot,\cdot\right)}$
matrix, and $\mathrm{GRU_{B}}\left(\cdot\right)$ is a regular, non-sparse
GRU used in place of the fully-connected layer with ReLU activation
in~(\ref{eq:WaveRNN}).

\subsection{Sampling from Probability Distribution}

Directly sampling from the output distribution can sometimes cause
excessive noise. This is addressed in~\cite{jin2018fftnet} by multiplying
the logits by a constant $c=2$ for voice sounds, which is equivalent
to lowering the ``temperature'' of the sampling process. Rather
than making a binary voicing decision, we instead set
\begin{equation}
c=1+\max\left(0,1.5g_{p}-0.5\right)\ ,\label{eq:sampling_temperature}
\end{equation}
where $g_{p}$ is the pitch correlation ($0<g_{p}<1$). As a second
step, we subtract a constant from the distribution to ensure that
any probability below that constant threshold $T$ becomes zero. This
prevents impulse noise caused by low probabilities. The modified probability
distribution becomes 
\begin{equation}
P'\left(e_{t}\right)=\mathcal{R}\left(\max\left[\mathcal{R}\left(\left[P\left(e_{t}\right)\right]^{c}\right)-T,0\right]\right)\,,\label{eq:sampling_floor}
\end{equation}
where the $\mathcal{R}\left(\cdot\right)$ operator renormalizes the
distribution to unity, both between the two steps and on the result.
We find that $T=0.002$ provides a good trade-off between reducing
impulse noise and preserving the naturalness of the speech. 

\subsection{Training Noise Injection}

\label{subsec:training-noise-injection}

When synthesizing speech, the network operates in conditions that
are different from those of the training because the generated samples
are different (more imperfect) than those used during training. This
mismatch can amplify and cause excessive distortion in the synthesis.
To make the network more robust to the mismatch, we add noise to the
input during training, as suggested in~\cite{jin2018fftnet}. 

\begin{figure}
\centering{\includegraphics[width=1\columnwidth]{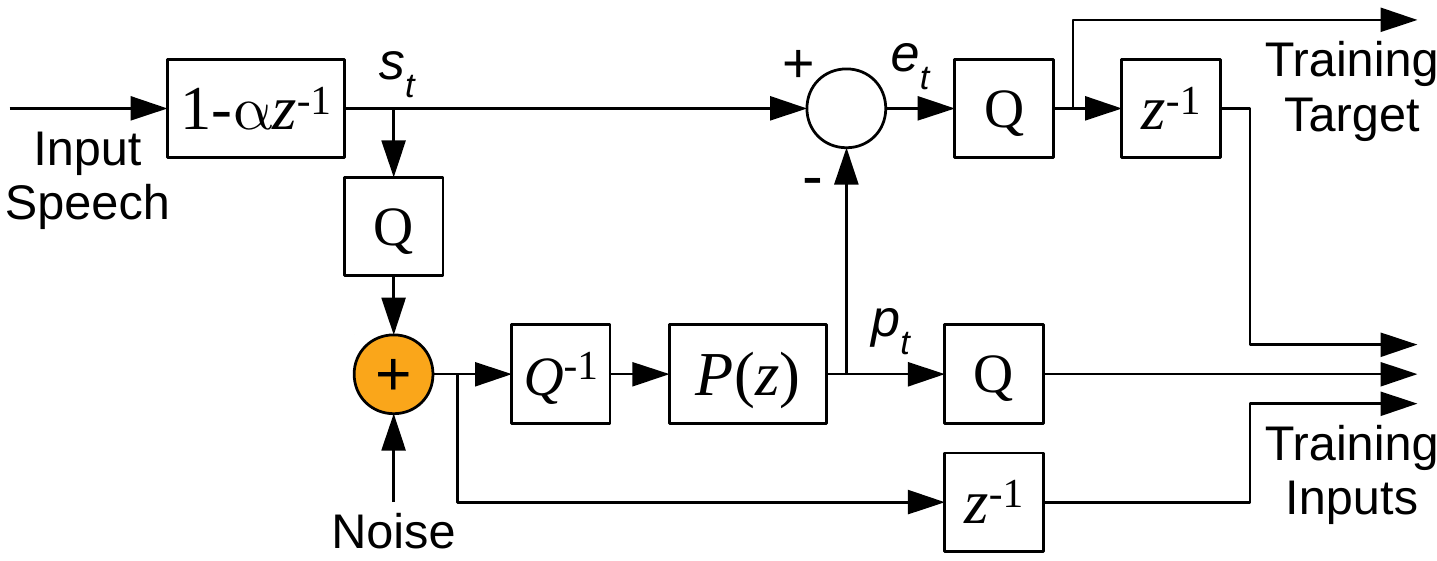}}

\caption{Noise injection during the training procedure, with $Q$ denoting
$\mu$-law quantization and $Q^{-1}$ denoting conversion from $\mu$-law
to linear. The prediction filter filter $P\left(z\right)=\sum_{k=1}^{M}a_{k}z^{-k}$
is applied to the noisy, quantized input. The excitation is computed
as the difference between the clean, unquantized input and the prediction.
Note that the noise is added in the $\mu$-law domain. \label{fig:Noise-injection}}
\end{figure}

The use of linear prediction makes the details of the noise injection
particularly important. When injecting noise in the signal, but training
the network on the clean excitation, we find that the system produces
artifacts similar to those of the pre-analysis-by-synthesis vocoder
era, where the noise has the same shape as the synthesis filter $\frac{1}{1-P\left(z\right)}$.
Instead, we find that by adding the noise as shown in Fig.~\ref{fig:Noise-injection},
the network effectively learns to minimize the error in the signal
domain because even though its output is the prediction residual,
one of its input is the same prediction that was used to compute that
residual. This is similar to the effect of analysis-by-synthesis in
CELP~\cite{atal1982new,schroeder1985code} and greatly reduces the
artifacts in the synthesized speech. 

To make the noise proportional to the signal amplitude, we inject
it directly in the $\mu$-law domain. We vary its distribution across
the training data from no noise to a uniform distribution in the $\left[-3,3\right]$
range.

\section{Evaluation}

\label{sec:Evaluation}

The source code for this work is available under an open-source license
at \url{https://github.com/mozilla/LPCNet/}. The evaluation in this
section is based on commit \texttt{0ddcda0}.

\subsection{Complexity}

The complexity of the proposed LPCNet model mostly comes from the
two GRUs as well as the dual fully-connected layer. It corresponds
to two operations (one add, one multiply) per weight, for each sample
produced and is given by
\begin{equation}
C=\left(3dN_{A}^{2}+3N_{B}\left(N_{A}+N_{B}\right)+2N_{B}Q\right)\cdot2F_{s}\ ,\label{eq:complexity}
\end{equation}
where $N_{A}$ and $N_{B}$ are the sizes of the two GRUs, $d$ is
the density of the sparse GRU, $Q$ is the number of $\mu$-law levels
and $F_{s}$ is the sampling rate. Using $N_{A}=384$, $N_{B}=16$
and $Q=256$ for wideband speech ($F_{s}=16000$), and considering
around 0.5~GFLOPS complexity for the neglected terms (biases, conditioning
network, activation functions, ...), we obtain a total complexity
around 2.8~GFLOPS. Real-time synthesis can be achieved on a single
core of an Apple A8 (iPhone~6) or with 20\% of a 2.4~GHz Intel Broadwell
core.

As a comparison, the speaker-dependent FFTNet model -- which claims
a lower complexity than the original WaveNet~\cite{van2016wavenet}
algorithm -- has a complexity around 16~GFLOPS~\cite{jin2018fftnet}.
The complexity of the original WaveRNN -- evaluated as a speaker-dependent
model -- is not explicitly stated, but our interpretation of the
data provided in the WaveRNN~\cite{kalchbrenner2018efficient} paper
suggests a complexity around 10~GFLOPS for the sparse, mobile version.
The complexity of SampleRNN is not explicitly stated either, but from
the paper, we estimate around 50~GFLOPS (mostly due to the 1024-unit
MLP layers).

\subsection{Experimental Setup}

\begin{figure}
\centering{\includegraphics[width=1\columnwidth]{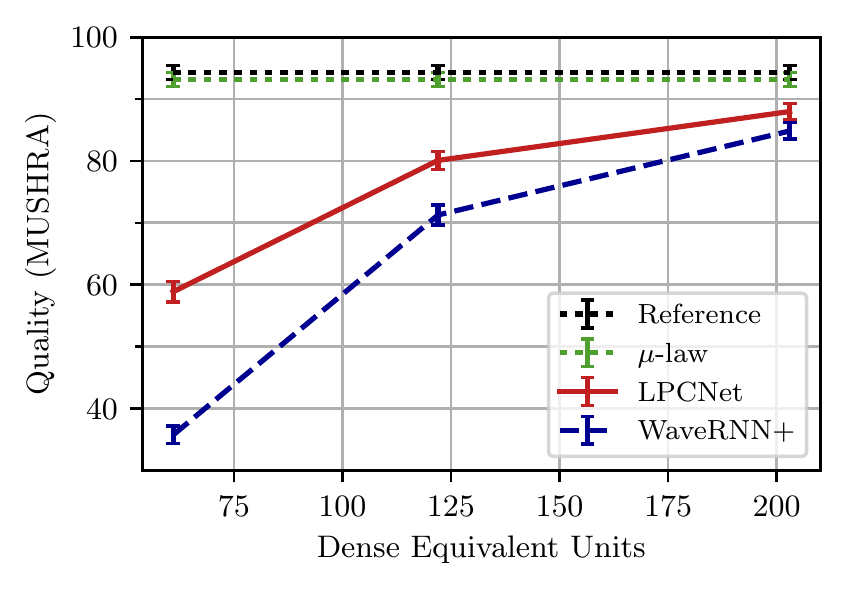}}

\caption{Subjective quality (MUSHRA) results as a function of the dense equivalent
number of units in $\mathrm{GRU_{A}}$.\label{fig:Subjective-quality-MUSHRA} }
\end{figure}

Although the proposed system can be either speaker-dependent or speaker-independent,
we evaluate it in the more challenging speaker-independent context.
To isolate the quality of the vocoder itself, we compute the features
directly from recorded speech samples. The cepstrum uses the same
band layout as~\cite{valin2017hybrid} and the pitch estimator is
based on an open-loop cross-correlation search.

The training data consists of only 4~hours of speech from the NTT
Multi-Lingual Speech Database for Telephonometry (21~languages),
from which we excluded all samples from the speakers used in testing.
Each network was trained for 120~epochs (230k~updates), with a batch
size size of 64, each sequence consisting of 15~10\nobreakdash-ms
frames. Training was performed on an Nvidia GPU with Keras\footnote{\url{https://keras.io/}}/Tensorflow\footnote{\url{https://www.tensorflow.org/}}
using the CuDNN GRU implementation and the AMSGrad~\cite{reddi2018convergence}
optimization method (Adam variant) with a step size $\alpha=\frac{\alpha_{0}}{1+\delta\cdot b}$
where $\alpha_{0}=0.001$, $\delta=5\times10^{-5},$ and $b$ is the
batch number. 

We compare LPCNet to an improved version of WaveRNN (denoted WaveRNN+)
that includes all of the improvements in Section~\ref{sec:LPCNet}
except for the LPC part, i.e. WaveRNN+ predicts sample $s_{t}$ only
from $s_{t-1}$ and the conditioning parameters. Each model is evaluated
with a main GRU of size $N_{A}$ equal to 192, 384, and 640~units,
and with a non-zero density $d=0.1$. These have the same number of
non-zero weights as equivalent dense GRUs of size 61, 122, and 203,
respectively (those GRU sizes match the ``equivalent'' sizes in~\cite{kalchbrenner2018efficient}).
In all cases, the size of the second GRU is $N_{B}=16$. For each
test sample, we compute the input features from the original audio
(ground truth), and then synthesize new audio with each of the models
under test. We also evaluate the effect of the $\mu$-law quantization
with pre-emphasis, as an upper-bound for the achievable quality.

\subsection{Quality Evaluation}

\label{subsec:Quality}

As reported in~\cite{kleijn2018wavenet}, objective quality metrics
such as PESQ and POLQA cannot adequately evaluate non-waveform, neural
vocoders. We conducted a subjective listening test with a MUSHRA-derived
methodology~\cite{BS1534}, where 8 utterances (2 male and 2 female
speakers) were each evaluated by 100 participants. The results in
Fig.~\ref{fig:Subjective-quality-MUSHRA} show that the quality of
LPCNet significantly exceeds that of WaveRNN+ at equal complexity.
Alternatively, it shows that the same quality is possible at a significantly
reduced complexity. The results also validate our assumption that
pre-emphasis makes the effect of $\mu$-law quantization noise negligible
compared to the synthesis artifacts. Being able to compute only a
single 256-value distribution also reduces complexity compared to
the original 16-bit WaveRNN.

The main audible artifact in the generated samples -- from both WaveRNN+
and LPCNet -- is some roughness due to noise between pitch harmonics.
One possible remedy -- which we did not yet investigate -- consists
of using post-denoising techniques, as suggested in~\cite{jin2018fftnet}.
A subset of the samples used in the listening test is available at
\url{https://people.xiph.org/~jm/demo/lpcnet/}.

\section{Conclusion}

\label{sec:Conclusion}

This work demonstrates that the efficiency of speaker-independent
speech synthesis can be improved by combining newer neural synthesis
techniques with the much older technique of linear prediction. In
addition to the main contribution of combining linear prediction with
WaveRNN, we make other contributions, such as the embedding of the
signal values, the improved sampling, as well as the pre-emphasis
prior to $\mu$-law quantization. We believe that the proposed model
is equally applicable to text-to-speech and to low-bitrate speech
coding.

Future work on LPCNet includes investigating whether long-term (pitch)
prediction can be incorporated as a way to further reduce the complexity.

\bibliographystyle{IEEEbib}
\bibliography{lpcnet}

\end{document}